# Understanding the Location of Resistance Change in the $Pr_{0.7}Ca_{0.3}MnO_3$ RRAM

Sandip Lashkare, Udayan Ganguly*

Department of Electrical Engineering, Indian Institute of Technology Bombay, Mumbai, India

**ABSTRACT:** $Pr_{1-x}Ca_xMnO_3$ (PCMO) based resistance random access memory (RRAM) is attractive in large scale memory and neuromorphic applications as it is non-filamentary, area scalable and has multiple resistance states along with excellent endurance and retention. The PCMO RRAM exhibit area scalable resistive switching when in contact with the reactive electrode. The interface redox reaction based resistance switching is observed electrically. Yet, whether resistance change occurs through *partial* (close to interface) or *entire* bulk is largely debated. Essentially, a two-terminal device is unable to provide direct evidence of the resistance change location in the PCMO RRAM. In this paper, we propose and experimentally demonstrate a novel three-terminal RRAM device in which a thin third terminal (~20nm) is inserted laterally in a typical vertical 2 terminal RRAM device of PCMO thickness of ~80nm. Using the 3T-RRAM method, we show that resistance change occurs largely at the upper bulk (near reactive electrode interface) - which is highly asymmetric. Yet it produces SCLC based resistance change with symmetric IV characteristics. It is the first time that an interface redox and bulk SCLC based resistance change has been experimentally shown as correlated and consistent - enabled by the 3rd terminal of the RRAM. Such a study enables a critical understanding of the device which enables the design and development of PCMO RRAM for memory and neuromorphic computing applications.

## INTRODUCTION

$Pr_{1-x}Ca_xMnO_3$(PCMO) RRAM has attracted widespread attention in crossbar memory and neuromorphic applications as it enables multiple non-volatile resistance states, forming less operation along with excellent endurance and retention.[1–5] Further, the area-scaling capability in PCMO enables high-density array required for large scale computation.

Chemically, the resistive switching has been attributed to a redox reaction with a reactive metal electrode (e.g. Ti or W) for metal oxide formation (dissolution) and oxygen vacancy poor (rich) region in the PCMO RRAM related to HRS (LRS)– by electrochemical migration of oxygen vacancies which is observed extensively by TEM analysis (Figure 1a,b).[6–10]

Electrically, a large number of studies demonstrate the SCLC current conduction mechanism, which is a bulk resistance change. Here, an oxygen-deficient *bulk* region of the PCMO film is presented as the cause of resistive switching (Figure 1c,d). The symmetric current-voltage characteristics on both the polarities and the current-voltage slope of 2 are demonstrated to supports the SCLC current conduction, which is a bulk phenomenon.[4,11-17]

The combination of *interface* driven redox resulting in *bulk* SCLC based current conduction requires further resolution. Sawa[18] and Asanuma[19] propose a bandgap enhancement by vacancies based on spectroscopic evidence. The vacancy an increase (decrease) which leads to increase (decrease) in Schottky barrier height is correlated to HRS (LRS). Alternatively, Herper[20] and Liu[21] have observed reactive metal oxide thickness modification which is physically correlated to HRS/LRS for $Ti/PCMO$ combination to propose open/short type mechanism. However, the interfacial oxide (here $WO_x$) may be highly non-stoichiometric and conducting.[22] Given that SCLC is electrically observed for both HRS and LRS and hence bulk limited, the interface $WO_x$ should not directly affect resistance. Yet the bandgap change in the *upper* bulk (below the reactive electrode) may affect the SCLC mechanism (Figure 1e,f). Further, the trap density may be modulated *entirely* across the bulk or *partially* limited to the upper bulk below the reactive electrode(Figure 1g,h). However, a major challenge is to resolve the location of the resistance change in a 2-terminal RRAM device.

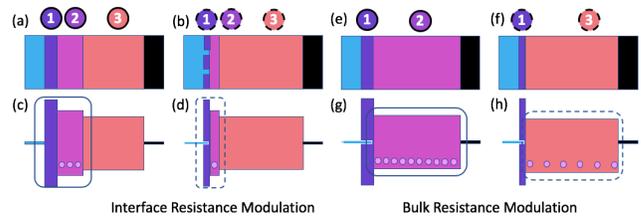

**Figure 1**. Schematic representation of resistive switching in PCMO sandwiched between reactive (light blue) and inert (black) electrodes. Electrochemical transport of oxygen ion produces (a) oxidized reactive electrode formation (region 1) and oxygen vacancy rich region in PCMO (region 2) for

HRS (b) partially dissolved metal oxide layer which supplies oxygen to fill up vacancies in PCMO for low resistance state (LRS). The vacancies modulation may be (a-b) partial or (c-d) be across the *entire* bulk. The resistance change may be due to (e-f) bandgap change in *partial* bulk or (g-h) trap density change across the *entire* bulk or a combination of the two effects. The resolution of the location of resistance change is a key measurement challenge.

Hence, in this paper, we propose and experimentally demonstrate a 3-terminal RRAM device in which a thin third terminal (~20nm) is inserted laterally in a typical vertical 2-terminal PCMO RRAM device. This 3rd terminal divides the PCMO film into two conduction regions – (1) upper bulk region (close to the reactive electrode) and (2) lower bulk region (close to the inert electrode). The independent measurements from different terminals show that not only the upper bulk region but the lower bulk region is also responsible for the resistive switching in the PCMO RRAM. Such a study details a critical understanding of the physical mechanism behind the resistive switching in the device and is essential for the design and development of memory and neuromorphic computing applications.

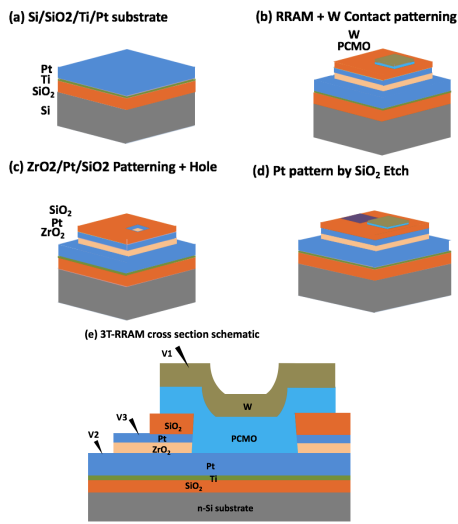

**Figure 2**. Three terminal RRAM (3T-RRAM) Process Flow. (a) First step, thermally grown SiO2 with Ti/Pt blanket deposition (T2) by sputtering, (b) second, ZrO2/Pt/SiO2 patterns with a hole in between using optical photolithography, (c) third, PCMO film deposition by second-level photolithography to fill the hole with subsequent top contact (W) deposition (T1), (d) Finally, wet etching of top SiO2 to open the Pt side contact (T3), (e) Cross-section schematic of the final 3T-RRAM structure.

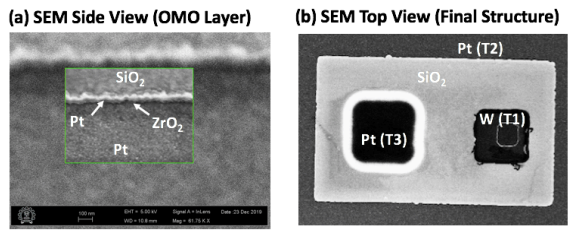

**Figure 3**. (a) SEM of OMO layer and (b) Top view SEM (3T-RRAM Final Structure)

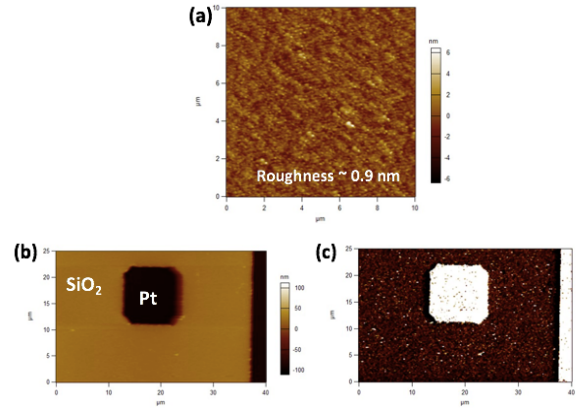

**Figure 4**. (a) AFM showing PCMO film roughness ~0.9nm, (b) AFM and (c) its phase image of OMO layer after step 2 (Figure 2) indicates two different surfaces (Pt and SiO$_2$).

## EXPERIMENTAL SECTION

The devices were fabricated on 4" Si <100> substrate. The substrate was thermally oxidized by rapid thermal oxidation at 1000°C to grow SiO$_2$ of 40 nm thickness. Thereafter, a Ti (25 nm)/Pt (150 nm) is deposited by sputtering in Ar ambient (Figure 2a). The platinum acts as bottom contact. Then ZrO2/Pt/SiO2 (OMO) layer of 30nm/20nm/50nm thickness is deposited with a hole in between by optical lithography and lift-off process (Figure 2b). The hole size defines the RRAM area which ranges from $5\mu m$ to $14\mu m$. The thin ~20nm Pt serves as side contact (3rd terminal). Then, Pr$_{1-x}$Ca$_x$MnO$_3$ (x=0.3) layer of ~80nm thickness is deposited by RF sputtering in Ar ambient at room temperature by second-level optical lithography filling hole created in the first level and lift-off process. The stack is then annealed at $750°C$ in $N_2$ ambient for 30s to crystallize PCMO film. Next, tungsten (W) top contact pads were created by photolithography and lift-off process (Figure 2c). Finally, top SiO2 is etched using BHF to access the side Pt contact by optical lithography (Figure 2d). The final cross-section 3-terminal RRAM (3T-RRAM) device schematic of the device is shown in Figure 2e showing W top contact (T1), Pt bottom contact (T2) and Pt side contact (T3). Here, at the bottom, ZrO$_2$ is used to isolate T3 and T2, as it is compatible with PCMO.[23] At the top, SiO$_2$ is used to enable T3 by wet etch and isolate T3 from T1.

The scanning electron microscope (SEM) at 45° angle shows the OMO layer whereas top view SEM shows the fabricated final device (Figure 3). Atomic force microscope (AFM) and its phase plot of the OMO layer with a hole between show the separation of the OMO layer and bottom Pt layer (Figure 4a). AFM and the phase plot shows the roughness of the PCMO film of ~0.9nm (Figure 4b,c).

The I-V measurements were carried out using Agilent B1500A/B1530A Semiconductor Parameter Analyzer.

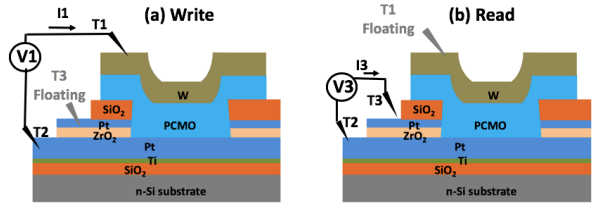

**Figure 5**. (a) Write scheme showing the applied voltage between T1 and T2 while T3 is kept floating, and (b) Read scheme showing the applied voltage between T3 and T2 while T1 is kept floating.

## DEVICE OPERATION

The write and read scheme of the PCMO RRAM have been utilized to separate the different conduction paths (Figure 5). For the write process, the bias is applied between T1 and T2, which forms a typical two-terminal PCMO RRAM with T3 kept floating. The write process gives a change in current (resistance) between T1 and T2 (Figure 5a) which includes an upper bulk region (close to T1) as well as lower bulk region (close to T2). For the read process, a small bias is applied between T3 and T2 with T1 kept floating (Figure 5b). The read process incorporates the change in current due to only the lower bulk region of PCMO film as it isolates T1 (an reactive electrode of RRAM).

Now for the as-fabricated PCMO devices of different areas, the current is plotted for the bias between V1 and V2 at 1V. Initially, the device is in low resistance state and the current flowing through device shows ~ 1:1 ratio with the area (Figure 6). This effect indicates the current scaling with the area, a signature of the non-filamentary nature of the device.

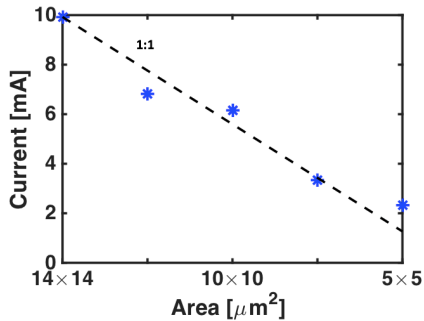

**Figure 6**. Current (Between T1 and T2) vs. Area of the PCMO devices showing the ~1:1 relation to the current and area indicating the current scaling with the area, a signature of the non-filamentary nature of the device.

The typical dciv characteristics of the PCMO RRAM is shown in Figure 7a using the write scheme. On the application of a small bias, the device does not change its state. The as-fabricated device is in a low resistance state. On positive polarity, after a threshold, the device changes its state from low resistance state (LRS) to a high resistance state (HRS). This is the RESET process. As a positive bias is applied, the current starts to rise. This increases the temperature in the device. After a threshold (Field and Temperature), the oxygen ions start to move from the PCMO film towards the W electrode leaving oxygen vacancies behind. The oxygen vacancies are the trap density in the device which in turn reduces the current flow creating a negative feedback loop. On further increasing the bias, the more number of oxygen vacancies are created further reducing the current and enabling high resistance state.[4]

On negative polarity, the device changes its state from HRS to LRS. This is the SET process. It can be observed that the SET is abrupt. The current starts to increase as the voltage is increased. The increase in current increases the Joule heating in the device. After a threshold (Field and Temperature), the oxygen ions start to move from the W electrode towards the device region filling the oxygen vacancies. The filling up of oxygen vacancies reduces the trap density in the device which in turn increases the current and further increase in the temperature. The increase in temperature further increases the current creating a positive feedback loop and hence an abrupt rise in current till compliance is reached.

In the low bias regime i.e. before resistive switching due to ionic motion, the IV characteristics can be described by $I \propto V^\alpha$ where an initial ohmic region ($\alpha = 1$) is followed by a space charge limited current (SCLC) region ($\alpha \approx 2$) (Figure 7b). The number of traps defines the current level of the device given by following equations,[24,25]

$$I_{Ohmic} \propto V \qquad (1)$$
$$I_{trapfree-SCLC} \propto V^2 \qquad (2)$$
$$I_{trap-SCLC} \propto \frac{I_{trapfree-SCLC}}{N_T} \qquad (3)$$

where $N_T$ is trap density in the bulk.

Here, a sequence of write and read is carried to observe the changes in the current in the different regions (upper bulk region, lower bulk region) of the device. The write current is shown in Figure 7a whereas the read current is shown in Figure 7c. First, the initial state of the device (as fabricated) is read by applying small bias V3 (= -0.5V). Then, the device is made to RESET (write) by V1 (= $V_{RESET}$) followed by a read. Finally, the device is made to SET (write) by V1 (= $V_{SET}$) followed by a read. The change in current (I1) while writing includes the changes occurring in the upper and lower bulk region of PCMO film. Whereas, the change in current (I3) while read occurs only due to the bulk PCMO region. It can be observed that after RESET the read current (I3) reduces whereas after the SET operation the read current increases. This increase and decrease in the read current due to indicate the resistance change occurring in the lower bulk region of the device as well.

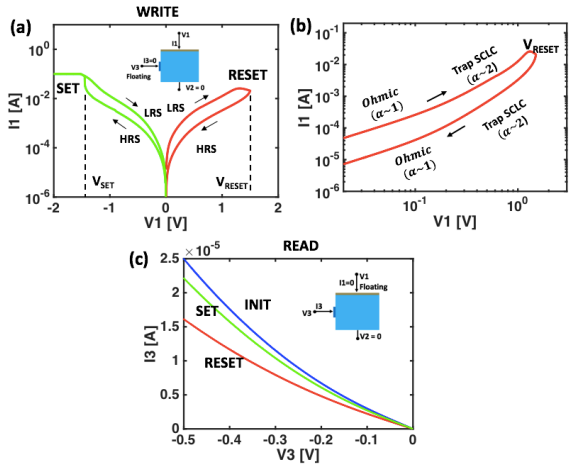

**Figure 7.** (a) WRITE: Typical dciv of the PCMO RRAM showing the SET and RESET on opposite polarities. The inset shows that the bias is applied between T1 and T2, (b) log-log dciv on the positive polarity showing the SCLC mechanism, (c) READ: V3 vs. I3 showing the change in the current which indicates the resistance change in the lower bulk region of the device.

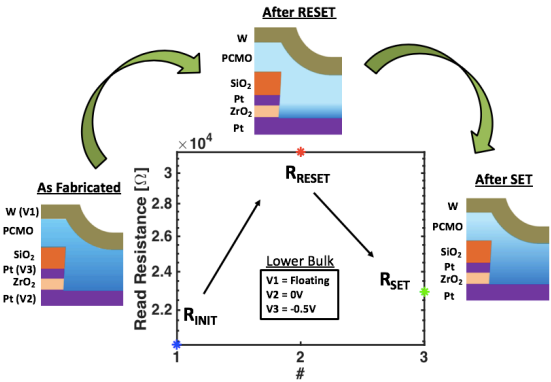

**Figure 8.** Read resistance of the PCMO device with a schematic showing the oxygen-rich or poor regions in the PCMO film after the application of SET and RESET bias.

## RESULTS AND DISCUSSION

As discussed earlier, the phenomenon of resistance change in the PCMO RRAM is due to the migration of oxygen ions towards (and away from) the reactive electrode. This forms an oxide layer on the metal electrode and an oxygen-deficient region in the PCMO film. As shown in Figure 8, in the first read, the device as fabricated in the LRS region. This shows the very less oxygen vacancies in the PCMO film. Then after RESET, the read resistance is increased. After RESET, the oxygen ions move towards the reactive electrode leaving oxygen vacancies behind. The increase in oxygen vacancies is the increase in the trap density and hence the increase in resistance. On SET, the oxygen ions move back into bulk and fill up the oxygen vacancies. This, in turn, is the reduction of trap density and hence the reduction in the resistance. It is observed that the resistance after SET does not go to the initial resistance of the device. This may be due to the oxygen ions staying trapped at the W/PCMO interface or the current is limited by series resistance (parasitic) and compliance (instrument limitation) and hence no further filling up of oxygen vacancies to come back to the as-fabricated resistance state.

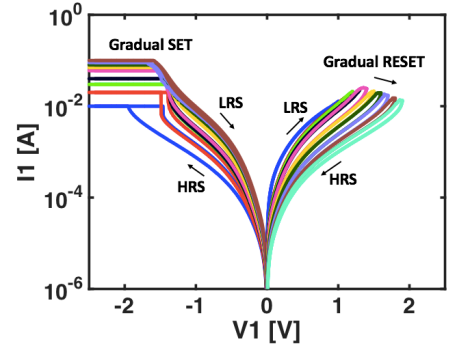

**Figure 9.** A Gradual change in resistance from LRS to HRS (by $V_{RESET}$) and from HRS to LRS (by current compliance).

To further observe the changes in resistance, the state of the device is changed gradually (Write process) and resistance is read for both upper + lower bulk region and only lower bulk region. The gradual RESET is obtained by the application of limiting $V_{RESET}$ whereas the gradual SET is obtained by limiting the current compliance (Figure 9).

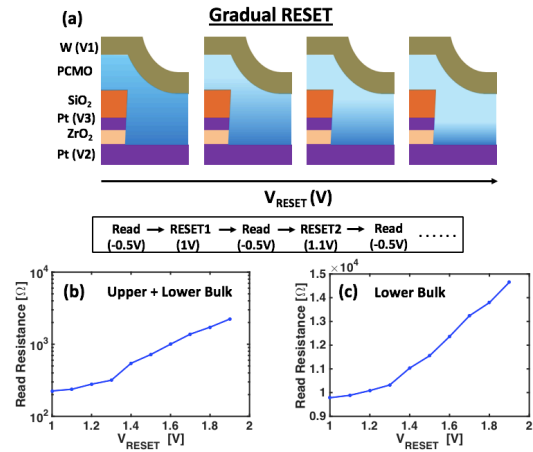

**Figure 10.** (a) The schematic showing gradual filling of the oxygen vacancies (shaded region) due to gradual RESET (by $V_{RESET}$, (b) gradual resistance increase in the upper + lower bulk region of the device with $V_{RESET}$ (c) gradual resistance change in the lower bulk region of the device $V_{RESET}$. A large change in the upper bulk region is observed owing to the higher trap density in the upper bulk region.

After every $V_{RESET}$, the current is read between both of the schemes i.e. between T1 and T2 to read the resistance change due to upper + lower bulk region and between T3 and T2 to read the resistance change only due to bulk region. It can be observed that the gradual resistance increase is occurring in both the cases for the increase in $V_{RESET}$ (Figure 10). The corresponding change in the device is shown by the increase in the shaded region (blue) indicating the increase in oxygen vacancies due to the migration of oxygen ions from the bulk region towards the interface.

Similar to RESET, the reverse trend i.e. decrease in resistance is observed in the gradual SET (Figure 11). The gradual SET was obtained by limiting the current compliance. The corresponding change in the device is shown by the decrease in the shaded region (blue) indicating the increase in oxygen vacancies due to the migration of oxygen ions from interface region towards the bulk region.

It can be seen that the resistance change in the upper + lower bulk region is high (10x) as compared to the change in only the lower bulk region (~1.5x). The large and small changes in resistance is due to a large number of oxygen vacancies (trap density) in the upper bulk and a small number of trap density in the lower bulk region. The large number trap density region in the PCMO film defines the maximum barrier for the SCLC conduction. The SCLC conduction is limited by maximum barrier given by the position of high trap density in the film but does not depend on the uniformity of trap density across the film. Here, the trap density is high in the upper bulk region and hence large resistance change is observed in the upper + lower bulk region read. However, as the lower bulk read does not have the high trap density region, a small resistance change is observed in the bulk region. Also, the bulk resistance observed is for lower area limited by T3 (20nm) as compared to T1 (5$\mu m$) and hence small resistance change.

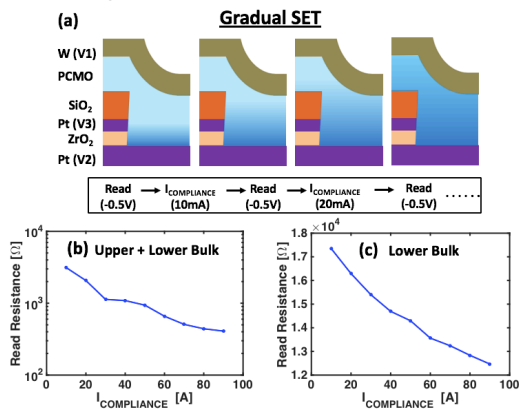

**Figure 11**. (a) The schematic showing gradual filling of the oxygen vacancies (shaded region) due to gradual SET (by current compliance), (b) gradual resistance decrease in the upper + lower bulk region of the device, (c) gradual resistance change of the lower bulk region of the device. A large change near the upper bulk region is observed owing to the higher trap density in the upper bulk region.

The resistance change is not observed from T1 to T3. This may be due to the majority of resistance change (due to T1 and T2) is centre region of PCMO than at the corner due to high field and hence no resistance change near the corner where the field is very low. Further, on the higher bias, no change in resistance switching is observed from T1(W) and T3 (Pt) even though it also forms a memory. This may be due to large current density confinement near T3 (non-reactive electrode) due to small area as opposed to the large area near the corner of T1 (reactive) due to convex shape or due to increased PCMO thickness by excess deposition at the corner.

## CONCLUSION

In this paper, to understand exact the resistance change location, we propose and demonstrate a three- terminal RRAM (3T-RRAM) in which a thin third terminal is inserted laterally in typical vertical 2-terminal PCMO RRAM device and is isolated from both of the RRAM electrodes. From this structure, we successfully resolve the upper and lower bulk conduction paths and show that not only the upper bulk region (close to the reactive electrode) but also the lower bulk region (close to the inert electrode) is responsible for the resistive switching in the PCMO RRAM. Such an understanding of the device is crucial to design and development of memory as well as neuromorphic computing applications.

## ACKNOWLEDGMENT

The work is partially funded by DST Nano Mission and Ministry of Electronics and IT (MeitY). It was performed at IIT Bombay Nanofab Facility. S. Lashkare is supported jointly by the Visvesvaraya PhD Scheme of MeitY, Government of India, being implemented by Digital India Corporation and Intel Ph.D. Fellowship.

## AUTHOR INFORMATION

**Corresponding Author**

* Email: udayan@ee.iitb.com